**Sensing via whole-body interactions with complex terrains: physical insights from robotic models**

Chen Li

Department of Mechanical Engineering, Johns Hopkins University, Baltimore, MD 21218, USA

## ABSTRACT

In nature, animals interact with complex terrains and often generate forces, deformations, and vibrations across the body and environment. Such interaction-induced mechanical information may guide behavior, yet its generation and use are difficult to resolve because movement rapidly and continually reconfigures the interactions that produce it. Sensory neurophysiology and biomechanics have revealed how mechanical information is transformed within sensory, neural, and musculoskeletal subsystems, whereas studies in complex terrains reveal rich whole-body interactions and behavioral outcomes. Connecting these levels remains a challenge. Here, I present a focused perspective on how biologically guided robotic physical models can help identify candidate mechanical information generated through whole-body interactions, test its physical mechanisms and behavioral benefits, and generate biological hypotheses. Three case studies illustrate this approach. In cluttered terrain, robot experiments guided by cockroach observations show that body–obstacle interaction forces can estimate potential energy landscape gradients, while exploratory movements can use this information to identify lower-resistance directions for locomotor transition to traverse. In uneven terrain, snake robot experiments show that contact feedback can maintain or regain propulsive contact during vertical bending, with distributed feedback enabling responses to contact changes along the body. In spider webs, robotic and biological experiments show that rapid leg crouching excites passive, partially decoupled vibrations whose frequencies carry information about prey presence and distance. Across these systems, correspondence between robots and animals supports the biological plausibility of these candidate physical mechanisms and motivates better measurements and models to determine how the relevant mechanical information is sensed and used to modulate behavior.

## INTRODUCTION

Sensing the external environment and responding appropriately are fundamental to animal behavior and locomotion (Cowan et al., 2014; Dickinson et al., 2000; Ijspeert and Daley, 2023; Mongeau et al., 2024; Nishikawa et al., 2007). Mechanosensing—the sensing of forces, contact, deformations, and vibrations generated within and outside the body—provides feedback for adaptive movement control, mediating reflexes, modulating locomotor rhythms, and contributing to action selection and state estimation (Dallmann et al., 2023; Goulding et al., 2025; Tuthill and Wilson, 2016). The mechanical information available to the nervous system is shaped not only by receptor and neural properties, but also by the mechanics of sensory structures, the musculoskeletal system, and interactions between the body and environment (Ijspeert and Daley, 2023; Nishikawa et al., 2007; Sane and McHenry, 2009). More broadly, active mechanosensing can shape sensory information through self-generated movements that alter animal–environment interactions (Bush et al., 2016; Hartmann, 2011; Prescott et al., 2011; Ritzmann et al., 2012; Staudacher et al., 2005). Thus, sensing and movement are coupled through both neural and mechanical processes, with sensory feedback, body mechanics, and environmental interaction jointly contributing to behavior (Cowan et al., 2014; Dickinson et al., 2000; Nishikawa et al., 2007).

Complex terrains make this coupling especially important, yet difficult to resolve. In uneven (Sponberg and Full, 2008), sparse or low-contact (Spagna et al., 2007), and densely cluttered (Li et al., 2015) terrains, on deformable and flowable substrates such as sand (Li et al., 2012; Li et al., 2013), and on animal-built structures such as spider webs (Mortimer, 2019), body–environment interactions can be distributed across many contacts, intermittent, and rapidly reconfigured by movement. For simplicity, hereafter we refer to these diverse environments as “complex terrains”. Contacts, forces, deformations, and vibrations can consequently change across the body and environment on timescales from milliseconds to seconds (Hung et al., 2026; Li et al., 2015; Mongeau et al., 2015; Sponberg and Full, 2008). Sensory neurophysiology and biomechanics have revealed how mechanical stimuli are transformed, often at the level of individual sensory, neural, or musculoskeletal subsystems and/or in relatively simple, controlled environmental contexts (Dallmann et al., 2023; Ijspeert and Daley, 2023; Nishikawa et al., 2007; Sane and McHenry, 2009; Tuthill and Wilson, 2016). Complementary studies of animals moving through complex terrains have revealed rich whole-body physical interactions and behavioral outcomes (Birn-Jeffery and Daley, 2012; Clifton et al., 2023; Fu et al., 2022; Harley et al., 2009; Kohlsdorf and Biewener, 2006; Li et al., 2015; Mongeau et al., 2012; Sponberg and Full, 2008). Connecting these levels remains an important challenge (Ijspeert and Daley, 2023; Sane and McHenry, 2009). Thus, for many complex terrain behaviors, a central challenge is to identify what mechanical information is generated and shaped by coupled whole-body–environment interactions, and how this information could regulate movement and behavior toward task-level outcomes. Such interaction-generated variables can also be difficult to measure simultaneously or manipulate independently in freely behaving animals.

Robotic physical models provide a complementary way to address this challenge because they permit simultaneous measurements and independent parameter variation that can be difficult or impossible in freely behaving animals (Aguilar et al., 2016; Aydin et al., 2019; Flammang and Porter, 2011; Full and Koditschek, 1999; Gravish and Lauder, 2018; Ijspeert, 2014; Ijspeert, 2020; Ijspeert and Daley, 2023; Koehl, 2003; Lauder, 2022; Webb, 2000; Webb, 2001). For complex terrains, robotic models of organisms can be combined with engineered terrains to enable systematic parameter variation and measurements of both the organism and the environment (Aguilar et al., 2016; Aydin et al., 2019; Othayoth et al., 2021), analogous to testing an airfoil in a wind tunnel.

Here, I review three case studies from my lab using this approach to examine how whole-body interactions with complex terrains generate candidate mechanical information that may support task-level behavioral outcomes. Across these cases, biologically guided robotic physical models help identify candidate mechanical information about whole-body–environment interactions difficult to obtain directly in living systems, test the potential behavioral benefits of using it, reveal the physical mechanisms, and generate targeted hypotheses for further biological investigation. This review provides a focused perspective, rather than a comprehensive survey of mechanosensing, locomotion, or robotics in complex terrains. Representative work from the broader field is discussed to place these case studies in context and identify open biological questions.

## CONSIDERATIONS OF ROBOTIC PHYSICAL MODELS

Robotic physical models can differ from animals in size, mass, morphology, material properties, compliance, sensing, actuation, and control. Achieving sufficient, question-relevant physical similarity—including dynamic similarity where appropriate (Alexander, 1991), not just morphological or kinematic resemblance—to the biological system is therefore a central concern (Flammang and Porter, 2011; Koehl, 2003; Lauder, 2022; Webb, 2001). However, in diverse, complex terrains, dynamic similarity can depend on which forces dominate rather than on a single universal scaling parameter (Alexander, 1991). This is because the relevant environmental interaction physics can combine contact geometry, friction, compliance, damping, gravity, intermittent collisions, flowability, heterogeneity, and/or stochasticity (Aguilar et al., 2016; Aydin et al., 2019; Othayoth et al., 2021). Establishing a simple scaling rule in a given complex terrain analogous to the Reynolds number may therefore require first establishing the dominant interaction physics. Moreover, it may be impossible or unnecessary to scale or match all aspects of the biological system simultaneously; instead, physical and robotic models can deliberately simplify the system and prioritize the features and interactions most relevant to the biological question, enabling controlled tests of hypotheses about organismal mechanics and control (Full and Koditschek, 1999;

Ijspeert, 2014; Koehl, 2003; Lauder, 2022; Webb, 2000). When the biological interaction is complex or little explored through robotic modeling, even a basic robotic model can create an experimentally tractable system and reveal which features are important (Aguilar et al., 2016; Gravish and Lauder, 2018; Ijspeert, 2020). Accordingly, in each case study, we follow the well-established practice: the robotic model is iteratively compared with its biological system and refined to reproduce the key biological features and interaction physics most relevant to the question (Flammang and Porter, 2011; Ijspeert, 2020; Ijspeert and Daley, 2023; Webb, 2001).

Robotic physical modeling complements theoretical and computational modeling approaches (Aguilar et al., 2016; Ijspeert and Daley, 2023; Koehl, 2003). Theoretical and computational models encode selected assumptions and parameters, enabling efficient exploration but making predictions dependent on modeling choices (Aguilar et al., 2016; Koehl, 2003). Robotic physical models instead realize physical interactions directly and can reveal emergent dynamics difficult to predict a priori, if sufficient question-relevant similarity is achieved (Gravish and Lauder, 2018; Ijspeert, 2020). Thus, they are particularly useful when the relevant interaction physics is unclear or difficult to model accurately. Although robots are themselves models, some dynamics emerge from real materials and contacts rather than being fully prescribed computationally, providing a complementary test of theoretical and computational models (Aguilar et al., 2016; Gravish and Lauder, 2018; Ijspeert, 2020; Koehl, 2003).

## BODY–OBSTACLE FORCE SENSING CAN HELP MULTI-LEGGED TRAVERSAL OF DENSE CLUTTER

Studies of insect locomotion have established that mechanosensory feedback from leg proprioceptors and load sensors helps regulate posture, stepping, and responses to perturbations (Büschges, 2005; Tuthill and Wilson, 2016; Zill et al., 2004). Antennae provide longer reach tactile information about the external environment: active exploration supports object localization and orientation (Okada and Toh, 2004; Okada and Toh, 2006), while contact location guides obstacle negotiation decisions in cockroaches (Harley et al., 2009; Ritzmann et al., 2012) and targeted leg placement during climbing in stick insects (Schütz and Dürr, 2011). In cockroaches, antennal feedback also supports rapid wall following and boundary exploration, while antenna–environment mechanics and body motion shape the information available for control (Cowan et al., 2006; Daltorio et al., 2015; Mongeau et al., 2013; Mongeau et al., 2014). Together, these studies show that tactile and proprioceptive information for locomotion is shaped by sensor mechanics, body movement, and environmental interaction.

Dense clutter traversal presents a distinct whole-body sensing problem: interactions with multiple large, flexible obstacles generate distributed forces that rapidly change with body configuration and may provide mechanical information for guiding locomotor transitions. We used forest floor-dwelling discoid cockroaches as our model organism, which is capable of diverse locomotor transitions through dense clutter (Li et al., 2015) (**Fig. 1A**). Our animal experiments across diverse obstacle types (Gart and Li, 2018; Gart et al., 2018; Han et al., 2021; Othayoth et al., 2021; Wang et al., 2022) (**Fig. 1B**) found that locomotion is not simply perturbed walking or running. Instead, cockroaches use whole-body interactions with the environment, undergo large 3-D body motions with limited kinetic energy and negligible momentum, and transition among distinct locomotor modes during traversal (**Fig. 1D**). We created minimalistic robotic physical models that reproduced these key biological features (**Fig. 1C**).

The physics of this cluttered obstacle traversal can be understood using a potential energy landscape approach. A potential energy landscape describes the gravitational and elastic potential energy of the self-propelled body–environment system as a function of body configurations relative to the environment (**Fig. 1E**). Stability basins of this landscape attract the animal's body into 3-D configurations that emerge as various locomotor modes (**Fig. 1E**). Thus, the animal must overcome the barriers separating these basins to make transitions toward traversal.

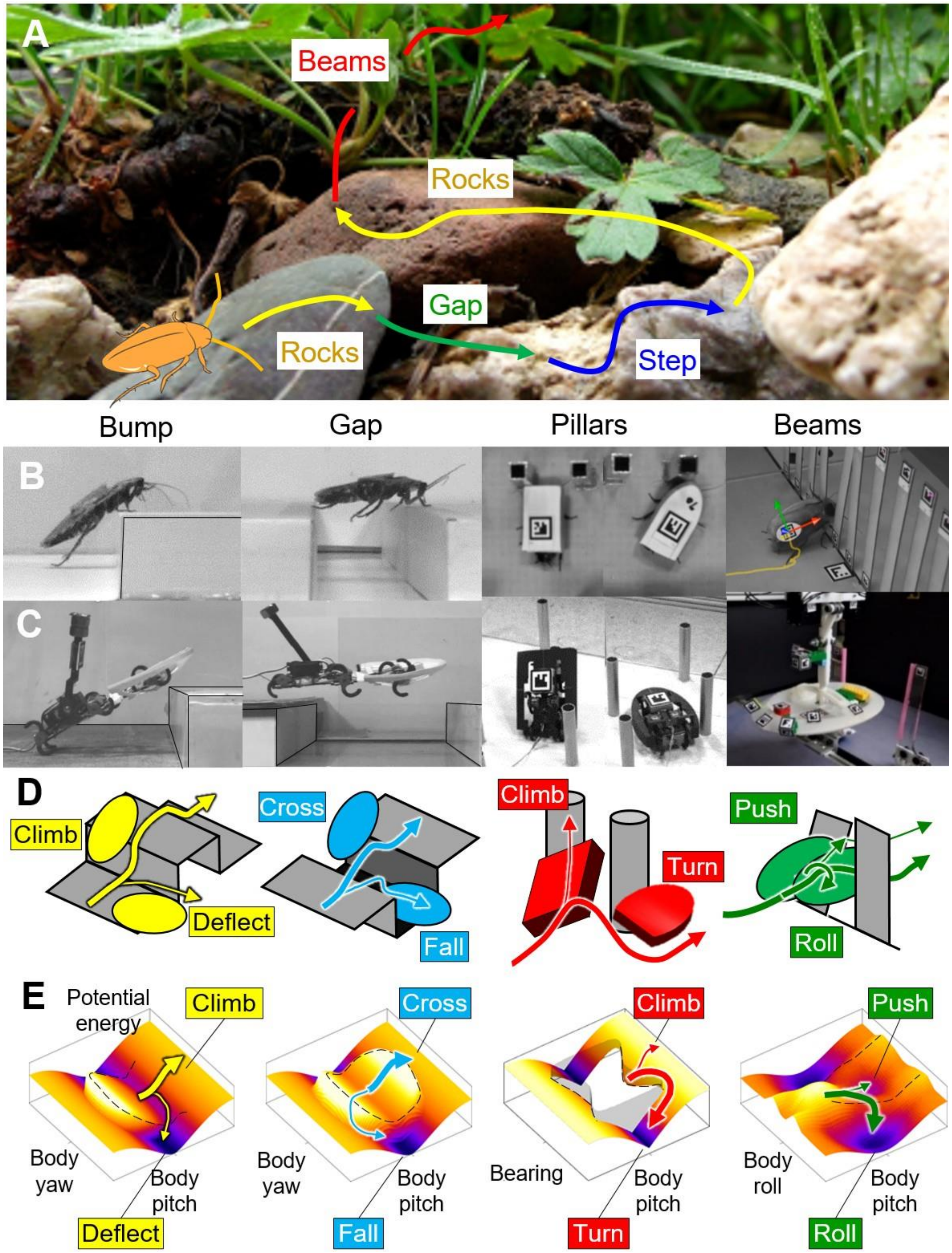

**Fig. 1. Multi-legged locomotor transitions to traverse dense clutter, and a potential energy landscape approach established through biological and robotic experiments. (A)** Discoid

cockroaches traverse dense clutter by using whole-body physical interactions with diverse types of obstacles to transition across distinct locomotor modes. **(B, C)** Systematic biological and robotic experiments of traversing representative obstacle types. **(D)** Locomotor modes (arrows). The ellipsoid represents the body, with each mode having large 3-D body rotation. **(E)** Potential energy landscapes with attractive stability basins. Attraction to each basin results in a distinct locomotor mode. Transitions between modes thus correspond with crossing barriers between basins (dashed curves). Adapted from (Othayoth et al., 2021).

Our focus system for studying mechanosensing here is the transition during traversal of cluttered grass-like beams, from a pitch mode in which it must push through the beams strenuously, to a roll mode in which it squeezes through narrow gaps more easily (**Fig. 2**). Remarkably, our animal observations (**Fig. 2A**) showed that in this transition, cockroaches cross the potential energy barrier near the saddle point, following transition routes of low resistance and energy cost (Wang, 2024) (**Fig. 2C**, iii', **Fig. 2D**, iii). Our feedforward robot experiments (**Fig. 2B**) found that sufficient kinetic energy fluctuation from oscillatory self-propulsion can help overcome barriers without sensory feedback control (Othayoth et al., 2020) (**Fig. 2E**, right). However, animal experiments showed that cockroaches can transition even when kinetic energy fluctuations are insufficient to overcome the barrier (Othayoth et al., 2020) (**Fig. 2E**, left), suggesting that sensory information may guide active adjustments that facilitate barrier crossing (Wang et al., 2022). Consistent with this possibility, cockroaches displayed substantial exploratory body motions and active leg adjustments during obstacle interaction (Wang et al., 2022).

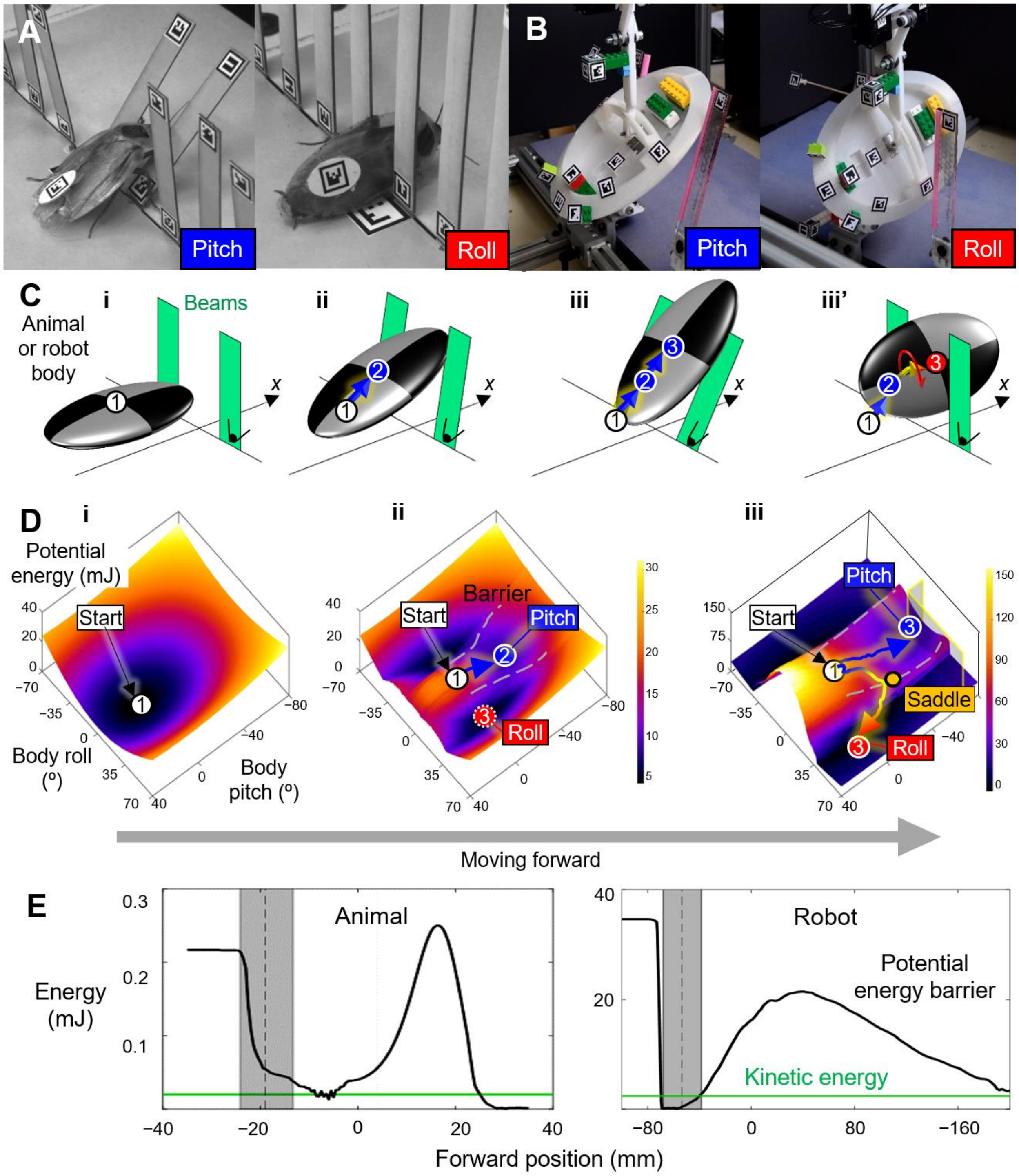

**Fig. 2. Model system of pitch-to-roll transition to traverse grass-like beams. (A, B)** Discoid cockroach (A) and feedforward robot (B) traversing beams using a pitch (left) or a roll (right) mode. (**C**) System configuration when the body is not in contact with the beams (i), interacts with beams in the pitch (ii, iii), or transitions to the roll mode (iii'). (**D**) Snapshots of potential energy landscape over body pitch–

roll space. Gray dashed curves show potential energy barriers separating the pitch and roll basins. Colored arrows show example system state trajectories. **(E)** Potential energy barrier from the pitch to the roll basin as a function of body forward position for the animal (left) and robot (right). Green line shows average kinetic energy fluctuation. Dashed line and gray band show mean ± s.d. of forward position at which transition occurs. Adapted from (Othayoth et al., 2020).

We further developed a simulation based on the minimalistic robot design to test how this mechanical information can guide transitions to traverse (Wang, 2024) (**Fig. 4**). The simulation showed that exploratory body motions (**Fig. 4C, D**) similar to those observed in cockroaches (**Fig. 4A, B**) can sample obstacle interaction forces to estimate potential energy gradients around the current body state, reveal local potential energy landscape structure, and identify directions toward saddle points (Wang, 2024). A saddle crossing controller then applies active forces to adjust the subsequent body motion in those directions, resulting in transitions resembling those observed in cockroaches (**Fig. 4D**).

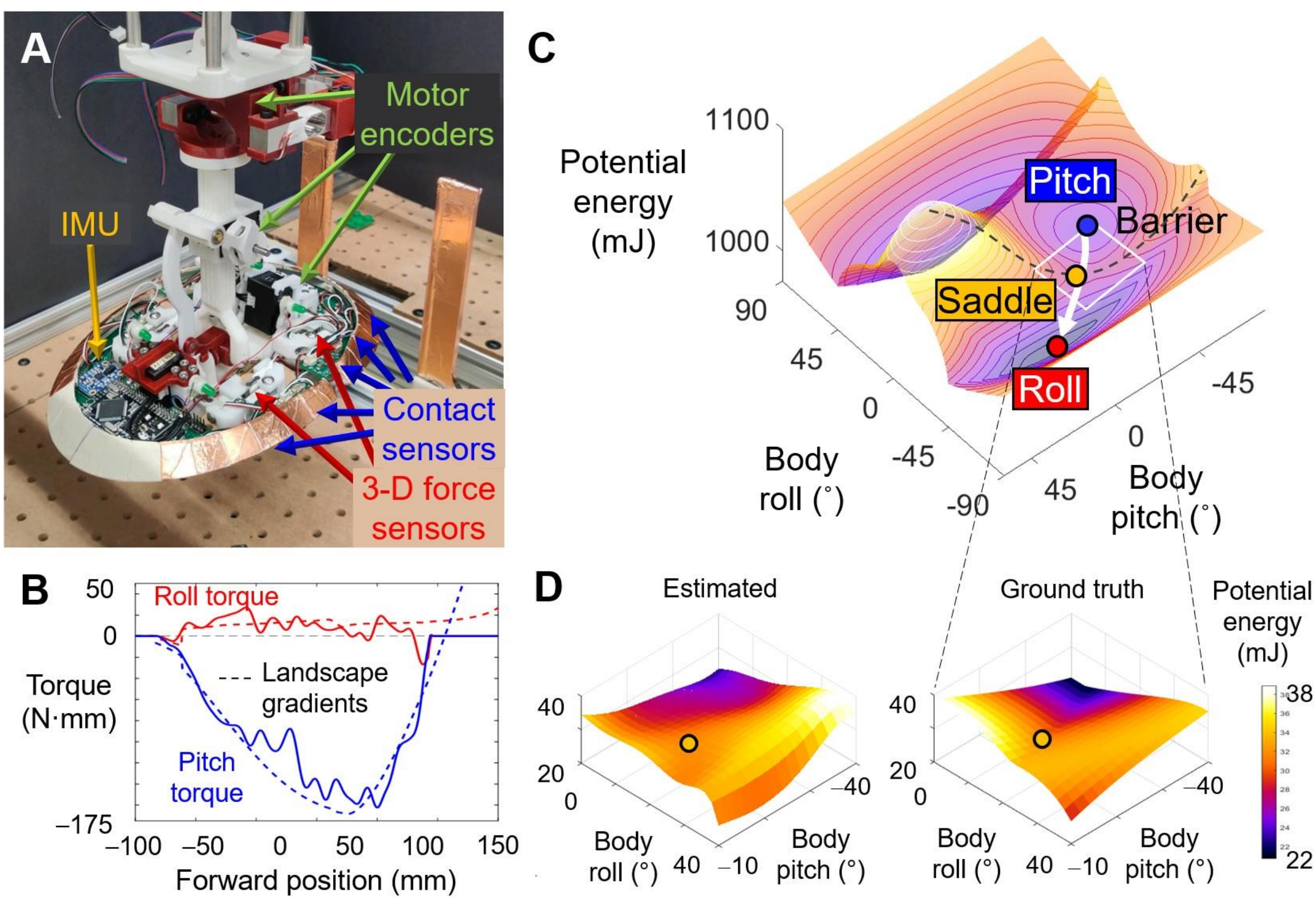

**Fig. 3. Obstacle contact force sensing provides a basis for potential energy landscape estimation. (A)** Robot with sensors for measuring obstacle contact forces and torques. **(B)** Representative torques (after subtracting Coulomb friction) along pitch and roll directions (solid curves) as a function of forward position for a representative body pitch and roll state, in comparison with negative gradients of the potential energy landscape (dashed curves). **(C)** A snapshot of the potential energy landscape over body pitch–roll space. Colored dots show the local minima of the pitch (blue) and roll (red) basins and the saddle (yellow) of the barrier between them (dashed curve). **(D)** Estimated potential energy landscape (left) compared with ground truth calculated from first principles (right). Landscape estimation performed over the white box in C. Adapted from (Wang et al., 2024).

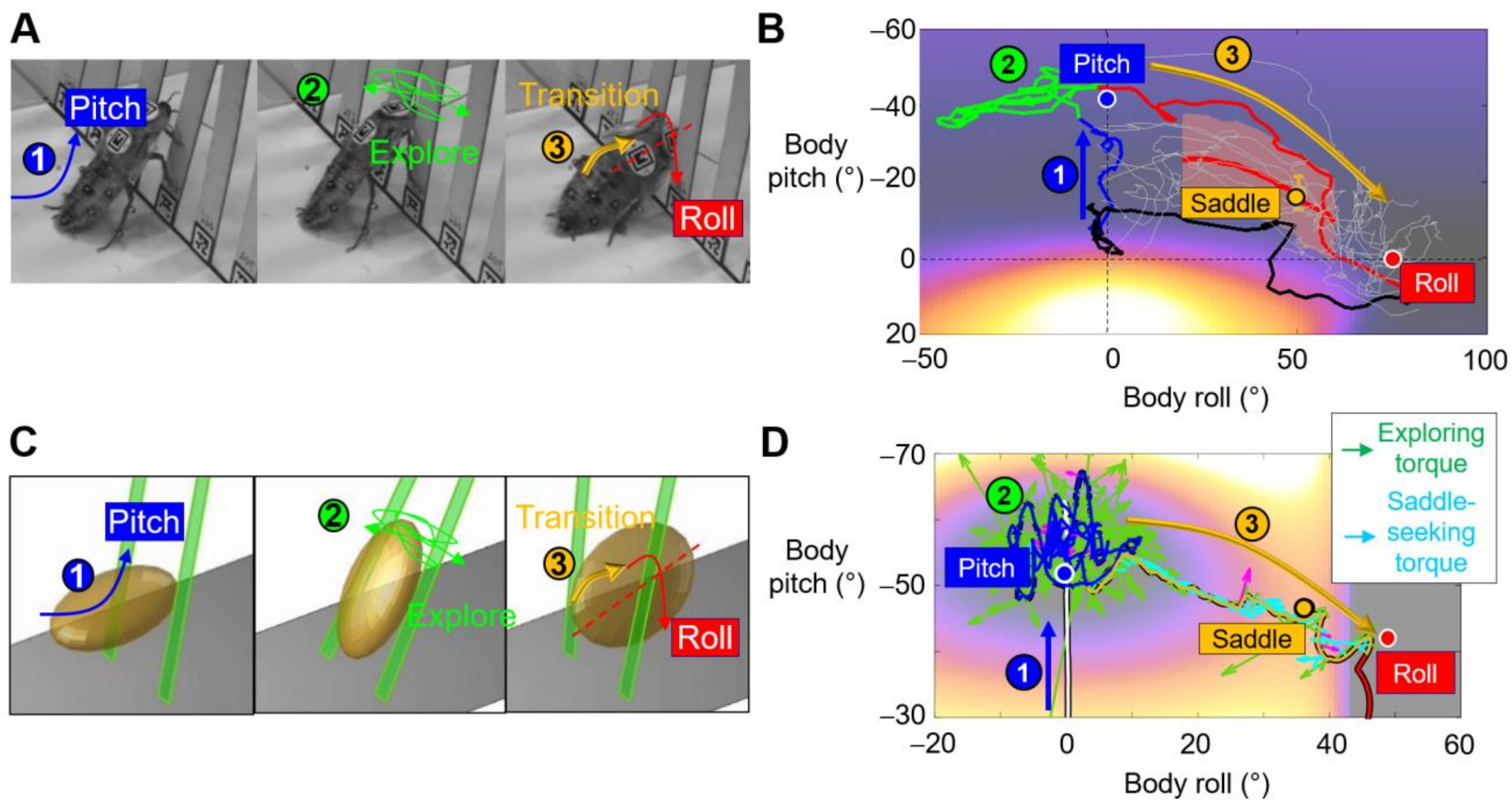

**Fig. 4. Bio-inspired control to traverse by detecting and crossing a saddle point with low resistance and energy cost. (A)** Discoid cockroaches often display exploratory body pitch and roll motions (middle) during transition from the pitch (left) to the roll (right) mode. **(B)** The cockroach's state trajectory on the potential energy landscape. **(C)** The simulated body displays similar exploratory body pitch and roll motions during pitch-to-roll transition. **(D)** Simulated system state trajectory on the potential energy

landscape. In (B, D), colored dots show the local minima of the pitch (blue) and roll (red) basins and the saddle (yellow) of the barrier between them. Adapted from (Wang, 2024).

The robot experiments and simulations do not demonstrate that cockroaches explicitly estimate potential energy gradients or use the same control algorithm. Rather, they identify a physical mechanism and motivate a biological hypothesis: coupled sensory, neural, musculoskeletal, and body–environment dynamics could use locally measured mechanical information to achieve functionally equivalent control without explicitly reconstructing a potential energy landscape. Studies in insects have established that mechanosensory feedback can provide information about local body configuration, loading, and contact relevant to locomotor regulation (Büschges, 2005; Tuthill and Wilson, 2016; Zill et al., 2004). Further simultaneous measurements and modeling of external obstacle forces, internal body loading, mechanosensory signals, and muscle activity during transitions can test which sensory signals encode information about local obstacle forces, how these signals are sampled during exploratory movements, and how they are processed by the nervous system to control the musculoskeletal system to generate body and leg movements that facilitate low-resistance transitions.

## DISTRIBUTED CONTACT SENSING CAN HELP LIMBLESS PROPULSION USING VERTICAL BENDING

Snake locomotion is a classic example of whole-body interaction with the environment. On flat, level surfaces, snakes can use lateral body bending to generate propulsion by exploiting frictional anisotropy (Hu et al., 2009). When discrete vertical asperities are available, generalist snakes can laterally bend their bodies to exploit them as propulsive push points (Gray and Lissmann, 1950; Moon and Gans, 1998; Schiebel et al., 2020). Robotic and modeling studies have shown that distributed local contact feedback along the body can modulate lateral body bending to exploit these push points to maintain propulsion (Kano and Ishiguro, 2020; Kano et al., 2012; Liljebäck et al., 2011; Liljebäck et al., 2012;

Sanfilippo et al., 2016). However, natural terrestrial environments are often uneven, providing terrain contacts above and below the body as well as laterally. In uneven terrain, distributed contacts and support conditions change along the body during movement, generating mechanical information that may help regulate 3-D body bending for propulsion.

We studied generalist corn snakes traversing uneven terrain composed of blocks with random height variation of up to five times body height (Fu et al., 2022). Using full-body reconstruction from discrete markers (Fu et al., 2021), we observed snakes rapidly traversing with smooth progression of the curved body and minimal lateral slipping, as if moving through a virtual tube (**Fig. 5A**). Planar projections identified segments likely to push laterally against terrain in the top view, and vertically in the flattened sagittal view (**Fig. 5B**). Snakes used these likely lateral and vertical push points with similar frequency (Fu et al., 2022). Thus, 3-D body bending can expand the terrain contacts available for propulsion, suggesting that studies emphasizing lateral bending on planar substrates capture only part of the propulsion strategies available in 3-D terrain.

However, control principles for lateral bending on level surfaces do not transfer directly to vertical bending in uneven terrain. During lateral bending, propulsion is generated largely by coordinating lateral forces from friction or vertical asperities in the horizontal plane. During vertical bending, body weight must also be supported and balanced against terrain contact forces: body regions can lose support, become suspended between contacts, or must conform to and push against sloped surfaces to generate forward propulsion rather than resistance. Thus, contact, loading, local terrain orientation, and body curvature become coupled through gravity, making appropriate body–terrain contact and body bending for propulsion particularly important (Fu and Li, 2023). Consistent with this challenge, snake robots using controlled compliance can conform to irregular 3-D terrain by yielding to contact forces but may become jammed or fail to move forward unless they selectively resist forces in some directions (Wang et al., 2020).

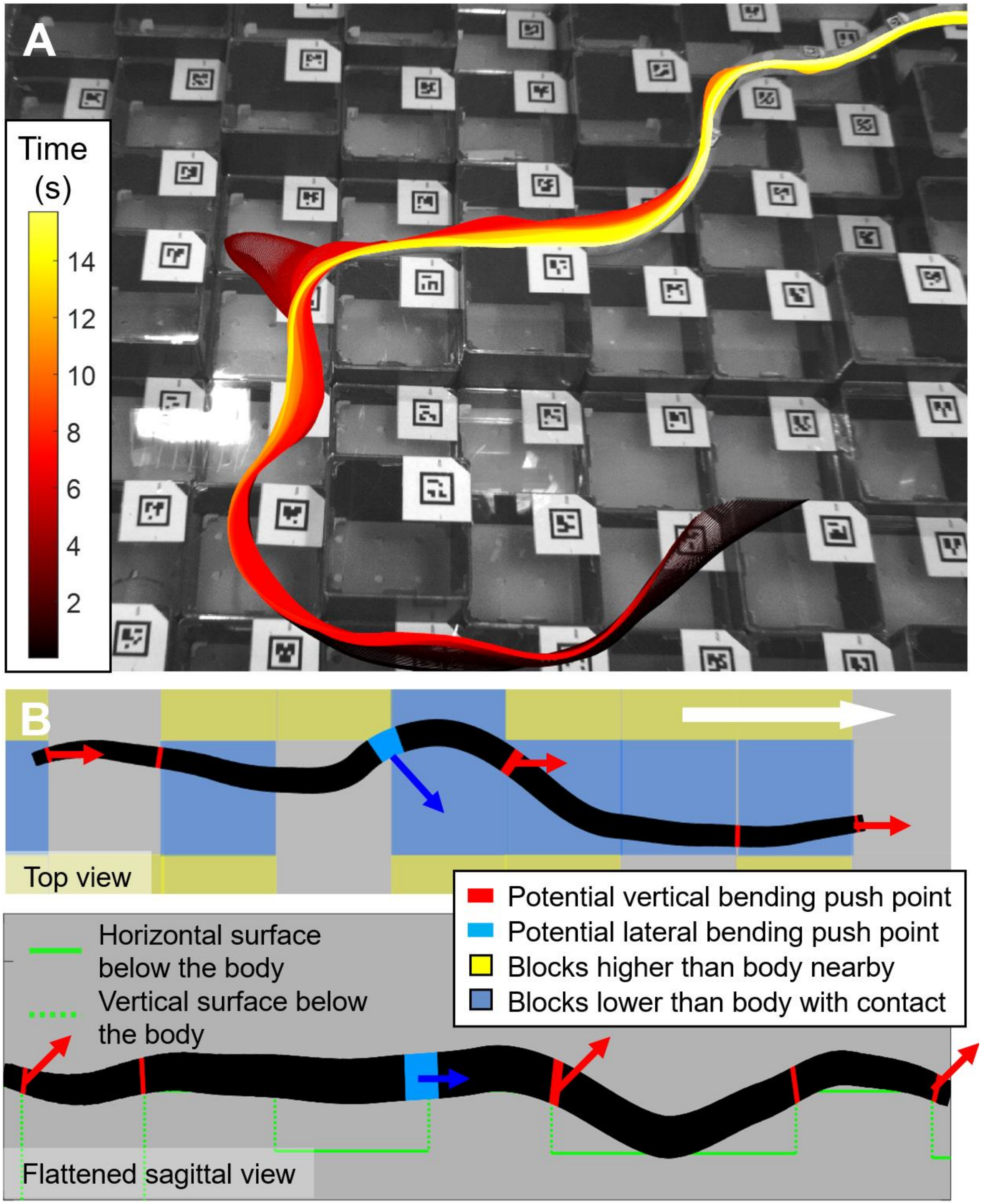

**Fig. 5. Snakes use 3-D body bending to propel through uneven terrain. (A)** Generalist corn snakes traverse uneven terrain with little lateral slipping, as if moving through a virtual tube. Colored curves show 3-D reconstruction of snake body over time. **(B)** A snapshot of the reconstructed snake body

interacting with terrain blocks, in top and flattened sagittal views (top to bottom). Snakes use potentially propulsive vertical and lateral contacts (red and blue regions and arrows) with similar frequency. Adapted from (Fu et al., 2022).

Snakes possess distributed tactile sensors (von Düring and Miller, 1979) and proprioceptive organs, including muscle and tendon stretch receptors (Crowe, 1992), but their use during freely moving 3-D locomotion remains difficult to measure. Consequently, much mechanistic work on contact feedback control in snake-like locomotion has used robotic or computational models, even for the simpler planar problem of exploiting vertical push points (Kano and Ishiguro, 2020; Kano et al., 2012; Liljebäck et al., 2011; Liljebäck et al., 2012; Sanfilippo et al., 2016). Therefore, we instrumented a snake robot with contact force sensing distributed along its body to test how contact feedback could regulate vertical body bending to generate and maintain propulsion (Fu and Li, 2023). As a tractable first step toward the full 3-D problem, we followed a biological study using a simplified terrain designed to elicit vertical bending propulsion while minimizing contributions from lateral bending (Jurestovsky et al., 2021) and tested whether and how the robot could benefit from feedback control using body–terrain contact information.

The robot was tested with feedforward control using prescribed motions without contact feedback, head-only feedback using contact information from the leading body region, and distributed feedback using local contact information from body segments along the robot (Fu and Li, 2023) (**Fig. 6**). Feedforward control generated propulsion when body bending matched terrain geometry but was vulnerable to perturbations that disrupted body–terrain contact (**Fig. 6A**). Head-only feedback could adapt when contact changes were sensed by the head (**Fig. 6B**), but it could not respond directly to perturbations occurring elsewhere along the body (**Fig. 6C**). With distributed feedback along the body, local contact information modulated body bending to maintain or regain propulsive terrain contact, enabling robust traversal under perturbations along the body (**Fig. 6D**). The robot’s motion resembled that of the snake (Jurestovsky et al., 2021) in that both used vertical bending to propel forward while conforming to terrain geometry, as if moving in a virtual tube.

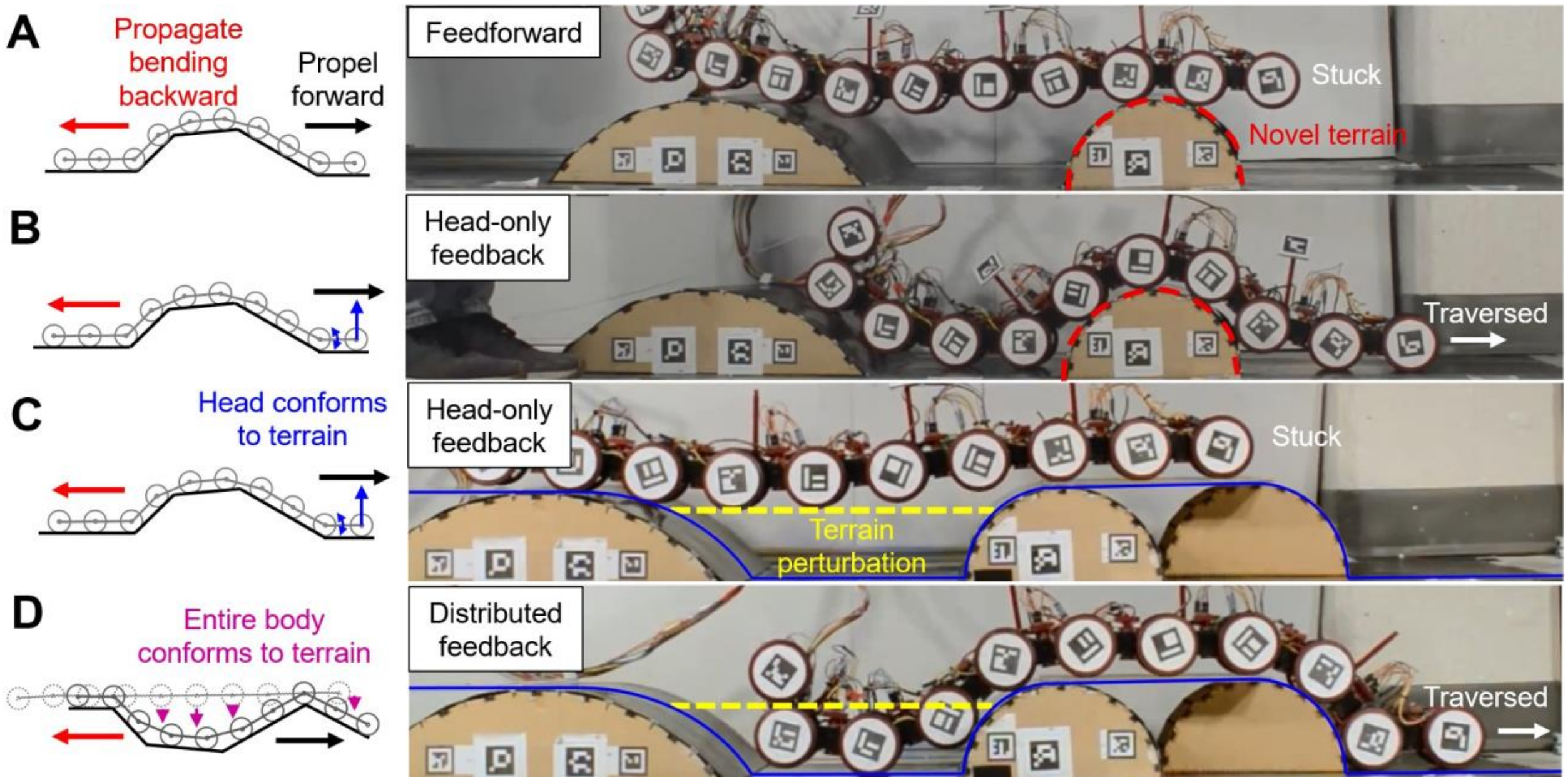

**Fig. 6. Snake robot traversing uneven terrain using vertical body bending with different control strategies. (A)** Feedforward control without sensory feedback**. (B, C)** Feedback control using contact sensing from the head only. **(D)** Feedback control using contact sensing from the entire body. Left: schematics of how each controller works. Right: snapshots of eventual outcomes. Red dashed curves in (A, B) show novel terrain encountered. Yellow dashed lines in (C, D) show where an initial terrain surface is suddenly removed to apply a perturbation. Adapted from (Fu and Li, 2023).

These robot experiments demonstrated that contact feedback increased the robustness of vertical bending propulsion, while distributed feedback enabled responses to changes in contact at various locations along the body. A simulation of a snake pushing against a wedge likewise found that force feedback can reduce fluctuations in forward motion and stabilize propulsion (Zhang et al., 2022). The robot experiments do not establish that snakes use the same sensors or neural circuitry. Rather, they identify a candidate physical mechanism by which local body–terrain contact information can regulate propulsion through vertical body bending as contact and support conditions change along the body. Further simultaneous measurements and modeling of ground reaction forces, muscle activity, and distributed tactile and proprioceptive signals can test which sensory signals encode local loading and how this information regulates vertical bending along the body to maintain propulsion in uneven terrain.

## RAPID LEG CROUCHING CAN HELP SPIDER VIBRATION SENSING OF PREY ON A WEB

Web-building spiders rely heavily on vibrations transmitted through the web to detect and localize prey (Barth, 2002; Foelix, 2011; Mortimer, 2019). The web is a predominantly planar, tensioned, and deformable animal-built structure, rather than a 3-D terrain. Nevertheless, this system presents similar challenges: spider behavior rapidly changes the whole-body–web–prey interactions and, consequently, the mechanical information available for sensing over milliseconds to seconds (Hung et al., 2026). Such mechanical information can in turn guide task-level prey capture behavior.

Research on spider vibration sensing has established that web structure, tension, silk mechanics, body posture, and receptor properties shape transmitted vibrations (Barth, 2002; Foelix, 2011; Landolfa and Barth, 1996; Masters, 1984a; Masters, 1984b; McConney et al., 2007; Mhatre et al., 2018; Mortimer, 2019; Mortimer et al., 2016; Mortimer et al., 2018; Mortimer et al., 2019; Wu et al., 2023). The web can therefore function as an extension of the spider's sensory system, with its architecture and mechanical properties shaping the information available to the animal (Japyassú and Laland, 2017; Mortimer, 2019). At longer timescales, spiders can actively construct and modify this sensory substrate through web-building behavior (Corver et al., 2021; Mortimer, 2019). Laser Doppler vibrometry can characterize spatial vibration transmission through measurements at multiple locations when stimuli and preparations are sufficiently repeatable, as demonstrated in spiders and other arthropods (Aicher et al., 1983; Dierkes and Barth, 1995; Mhatre et al., 2018; Sivalinghem, 2020; Stritih Peljhan and Strauß, 2018; Wu et al., 2023). Experiments and computational modeling have further shown how spider posture (treated as static during vibration measurements) and whole-body dynamics can mechanically filter vibrations before they reach sensory organs (Mhatre et al., 2018; Wu et al., 2023). These studies address questions related to, but distinct from, the robotic physical work reviewed here. A remaining experimental challenge is obtaining synchronized, distributed measurements across a coupled spider–web–prey system while both animals behave and rapidly alter the mechanical state of the system in milliseconds to seconds. Fine web threads can be difficult to keep in focus during laser Doppler vibrometry, and full-web laser Doppler vibrometry scans are not feasible on the timescales of rapid, non-repeatable behaviors. Consequently, the physical effects of rapid spider behavior on the coupled system remain less understood.

Our study followed animal behavioral observations from our collaborator's lab (Hung et al., 2026). During prey capture on its horizontal web, the orb-weaver spider *Uloborus diversus* alternates among static, rapid leg crouching, and other high velocity movement states. When prey-generated vibrations weaken, these spiders are more likely to rapidly crouch their legs, suggesting that rapid leg crouching may be an active sensing strategy.

As a first step to emulate this, we created a robotic physical model of a spider–web–prey system (**Fig. 7A**), with rapid leg crouching behavior relevant to sensing (**Fig. 7B, C**) but no full locomotion capabilities. The model consisted of a spider robot at the center of the web, a horizontal physical web, and a prey robot that could shake once, repeatedly, or stay inactive. The web design followed a previous vertical physical web for studying web vibration propagation (Otto et al., 2018). Due to manufacturing constraints, we could not match some physical properties such as the mass ratios of the web to the spider and prey robots or the robots' numbers of degrees of freedom to those in the biological system. The physical web was also more highly tensioned and sagged less than the *U. diversus* web, and it had substantially fewer radial threads. Nevertheless, the robotic system reproduced key qualitative features relevant to the question—low-frequency, large-amplitude transverse and lateral vibrations during and after rapid leg crouching and prey active shaking, as also observed in the living system (Hung et al., 2026; Lin et al., 2025). Such motion of thin web threads can be difficult to track continuously using laser Doppler vibrometry (Otto, 2019). This robotic system enabled synchronized measurement of 3-D web vibrations, spider robot leg vibrations, radial thread tensions, and actuator motions, together with systematic variation of spider and prey robot behaviors, which were challenging to achieve in the biological system.

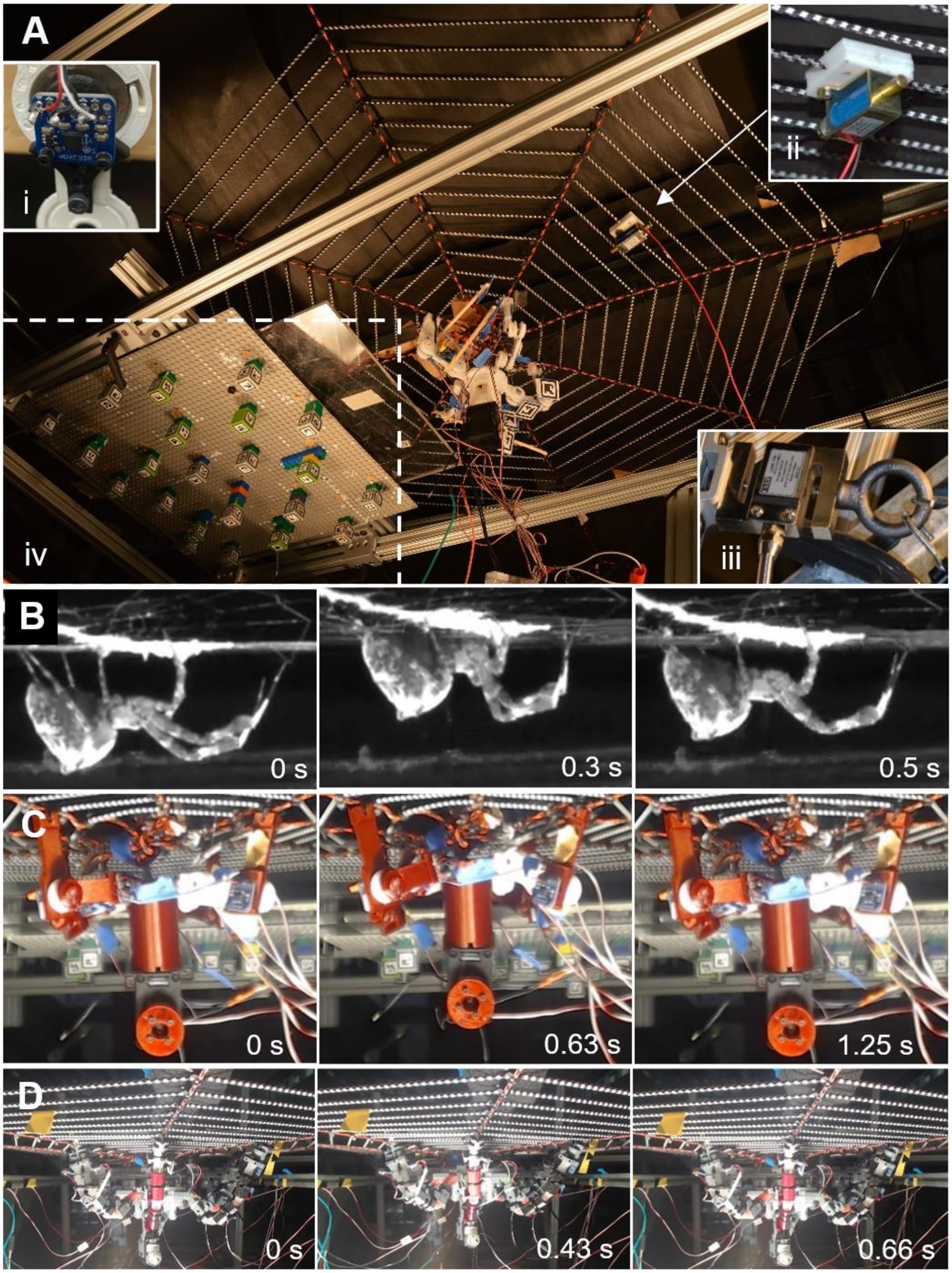

**Fig. 7. Robophysical modeling of how spiders use rapid leg behaviors to enhance vibration sensing of prey on web. (A)** Robophysical model of a spider–web–prey system. Insets (i—iv) show the spider robot leg accelerometer, prey robot, load cell at the end of each radial thread, and 3-D calibration object used to measure 3-D web vibrations from high-speed video tracking of web stripes. **(B)** Snapshots of rapid leg crouching by a *Uloborus diversus* spider on the web during prey sensing. **(C, D)** Snapshots of rapid leg crouching by the simplified (C) and refined (D) spider robots on the web. Video shown in B: courtesy of Hsin-Yi Hung from Andrew Gordus lab. A–C adapted from (Lin et al., 2025). D adapted from (Sun et al., 2026).

We performed preliminary robot experiments varying both spider robot and prey robot conditions guided by biological observations. Informed by these experiments, we focused first on the case in which the spider robot rapidly crouches its legs once and then senses the subsequent passive leg vibrations, while the prey robot condition is varied. This allowed us to isolate the physical consequences of a rapid leg crouch (Lin et al., 2025). We found that this excites large passive vibrations of the coupled web–spider–prey system that can reveal prey robot presence and distance (**Fig. 8**). Without a prey robot, the vibrations sensed by the spider robot after rapid leg crouching exhibit one dominant frequency peak (**Fig. 8A**, black). With a prey robot present, the first peak shifts slightly, and a second frequency peak emerges (**Fig. 8A**, blue, red), regardless of whether the prey robot shakes. Spatial profiles of web vibrations showed that the first peak reflects passive vibration of the spider robot and most of the web, whereas the second peak reflects passive vibration of the prey robot on its spiral thread (**Fig. 8B**). The frequency of the second peak increases as the prey robot becomes closer to the spider robot at the web center (**Fig. 8A**, blue to red), because its thread becomes shorter (**Fig. 8A**), analogous to the higher pitch of a shorter guitar string. This result is consistent with previous physical modeling of spider web vibration propagation using a static spider physical model on a vertical web, with vibrations generated by experimenter plucking (Otto et al., 2018). Guided by these robotic results, we analyzed the living spider–web–prey system and found evidence of the same two-frequency phenomenon, including passive prey vibrations at a higher frequency than those of the spider and most of the web (**Fig. 8C**). This provided biological support that, despite their substantial differences, our robotic physical model captures important features of the living system.

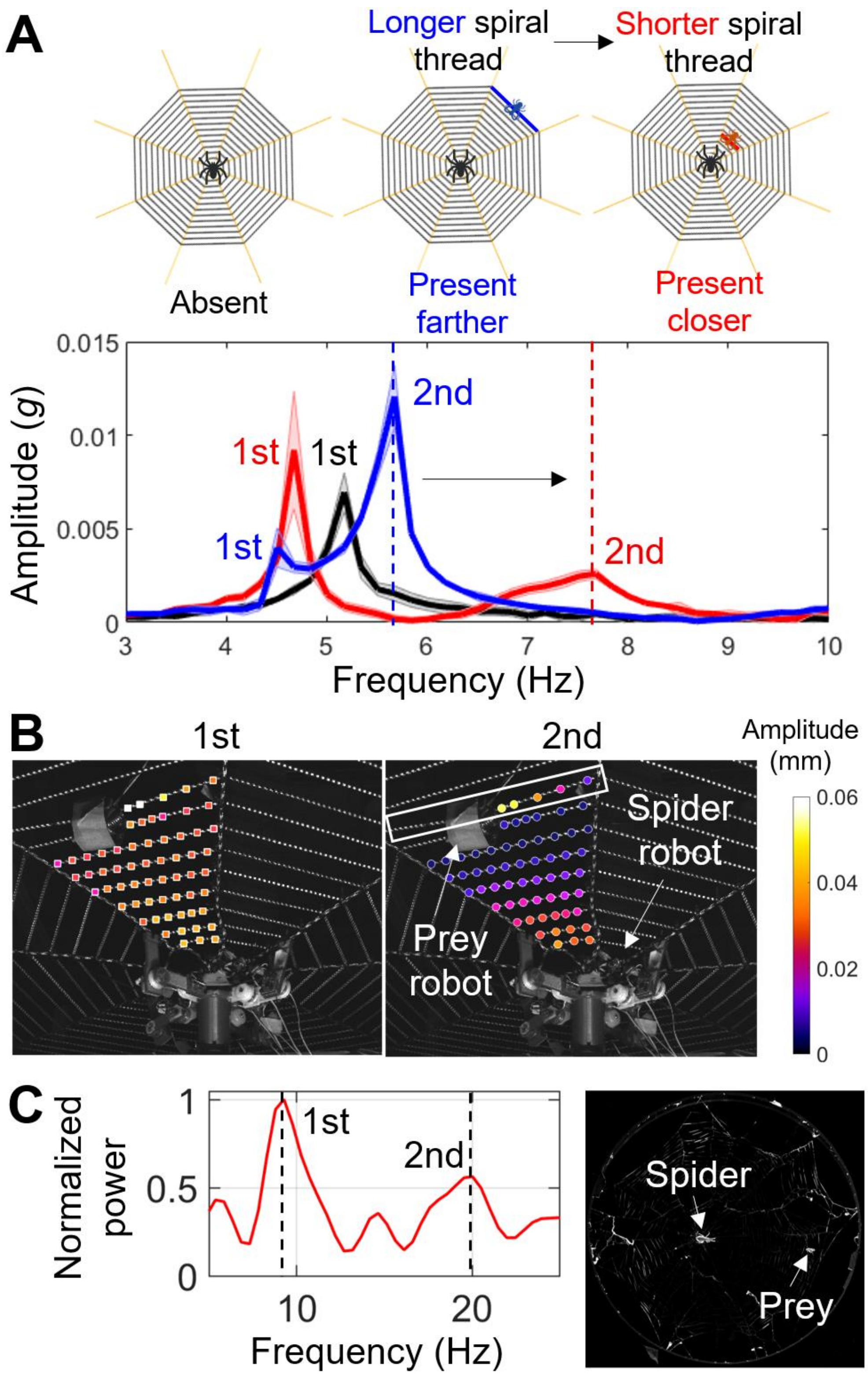

**Fig. 8. Rapid leg crouching of the spider robot helps detect the presence and distance of the prey robot. (A)** Fourier transforms of passive vibrations sensed by spider robot legs after a rapid leg crouch, for three conditions: prey robot absent (black), prey robot present farther from the spider robot (blue), and prey robot present closer to the spider robot (red). **(B)** Spatial profiles of transverse (out-of-plane) web vibration amplitudes for passive vibrations at the first and second frequency peaks identified when the prey robot is present. **(C)** Fourier transform of passive vibrations of the web near the prey after a rapid crouch of the spider in the biological system (inset). Videos to generate C: courtesy of Hsin-Yi Hung from Andrew Gordus lab. Adapted from (Lin et al., 2025).

The robotic physical model identifies a physical mechanism by which rapid leg crouching can function as an active sensing behavior: a rapid leg crouch excites the system to vibrate passively as two partially decoupled subsystems with distinct natural frequencies, making prey-related vibrations mechanically distinguishable from vibrations of the spider robot and the rest of the web. We recently created a refined robotic physical model that better approximates the living system with eight legs, more spider-like leg morphology and joint structure, and a greater rapid leg crouching range; experiments reproduced the key vibration features observed with the original robotic model (Sun et al., 2026). We also recently built a physical web that better resembles *U. diversus* web with denser radial threads and larger sagging through lower tension, and early experiments also reproduced similar key features.

Our robotic physical modeling, together with biological observations (Hung et al., 2026), suggested that low-frequency, large-amplitude vibrations, previously regarded as noise (Eberhard, 2020), may carry useful prey-related information for spider sensing on webs. This is consistent with evidence that low-frequency web vibrations can elicit prey capture responses in web-building spiders (Klärner and Barth, 1982) and that prey caught on spider webs often themselves generate vibrations in this low-frequency range (Hung et al., 2026; Landolfa and Barth, 1996; Masters, 1984b; Mortimer, 2019). Biological experiments that exploit natural variation in spider leg behavior, prey distance, spider and prey mass, and web geometry and tension, combined with synchronized measurements and modeling of web and leg vibrations, sensory responses, and internal state transitions can test which vibration features induced by rapid leg crouching are encoded by the spider and how this information is transformed into prey capture behavior. Advances in measurement techniques for capturing large-amplitude vibrations throughout the system with rapidly behaving animals, and in dynamic simulations that resolve such rapid behaviors, will be required to advance this area.

## CROSS-SYSTEM SYNTHESIS

The three systems differ in morphology, environment, and behavioral function, but share a common systems-level problem: rapidly changing whole-body–environment interactions that generate mechanical information relevant to task-level outcomes. These interactions are challenging to study in freely behaving animals, because the rapidly changing and distributed contacts, forces, deformations, and vibrations cannot easily be measured synchronously or manipulated independently.

Guided by biological observations, robotic physical models can reproduce salient behaviors and body–environment interaction physics in controlled, instrumented systems. This enables systematic measurement and perturbation to reveal candidate mechanical information, its potential benefits, and the physical mechanism that generates and uses it, before the biological implementation is known. By simplifying these systems while retaining question-relevant interaction physics, robotic models can reduce high-dimensional interaction problems to a smaller set of informative mechanical variables that can guide behavior with relatively simple rules. They also help identify the dominant interaction physics in such complex systems, which may inform further analysis of dynamic similarity.

The three case studies illustrate three levels of evidence linking robotic physical mechanisms to animal behavior. At the first level, robotic models identify candidate mechanical information generated through whole-body–environment interaction. In cluttered terrain, body–obstacle interaction forces provide information about local resistance to locomotor transitions. In uneven terrain, local body–terrain contact information provides information about where and how the body is supported, which can be used to regulate vertical bending for propulsion. In web vibration sensing, distinct passive vibration frequencies following rapid leg crouching provide information about prey presence and distance.

At the second level, robot experiments test how the identified information can be used to achieve a task. Cockroach robot experiments showed that body–obstacle interaction forces can estimate local potential energy gradients, and simulations showed how exploratory movements while sensing these forces can reveal local landscape structure and guide lower-resistance transitions. Snake robot experiments showed that contact feedback can regulate local vertical bending to maintain or regain propulsive contact as support changes along the body. Spider robot experiments showed that a rapid leg crouch excites passive prey-associated vibrations at a distinct frequency that reduces with supporting thread length, revealing prey robot presence and distance.

At the third level, correspondence between animal and robotic physical systems strengthens the biological plausibility of these mechanisms. In cockroaches and cockroach robots, traversal involves large 3-D body motions, limited kinetic energy, and transitions among similar locomotor modes through similar obstacles. The robot and simulation show that locally measured obstacle forces can provide information for identifying lower-resistance transition directions toward which to move, when kinetic energy fluctuations alone are insufficient for barrier crossing, as observed in the animal. In snakes and snake robots, vertical bending enables the body to conform to uneven terrain and generate forward progression with little lateral slip, as if moving through a virtual tube. The robot reproduces this characteristic whole-body motion, and contact feedback restores propulsive contact when perturbations disrupt support, consistent with the robust progression observed in snakes. This correspondence supports the plausibility that snakes similarly use local contact information along the body to regulate vertical bending and maintain propulsion in uneven terrain. In both living and robotic spider–web–prey systems, rapid leg crouching produces large web motions, with similar prey-associated vibration features. These large-amplitude, low-frequency vibrations of the coupled spider–web–prey system occupy a regime that is difficult to characterize with conventional laser Doppler vibrometry and has therefore been less explored. The correspondence between the robotic and living systems thus points to potentially new mechanical information available for prey sensing during rapid behavior. In all three cases, the robotic model reproduces the salient body movements, environmental interactions, and behavioral outcomes observed in the animal. Because the robot achieves the shared outcome through the candidate physical mechanism using the mechanical information identified above, the physical mechanism and information may together play a similar behavioral role in the animal, even though the specific biological implementation is not yet known.

Testing these biological hypotheses and determining the biological implementation will require further development of experimental and modeling techniques suited to these less commonly used model organisms during freely behaving interactions with complex environments. For cockroaches and snakes, important advances include miniaturized and distributed measurements of contact forces, body loading, muscle activity, and sensory signals that can be used during contact-rich 3-D movement. Advances in full-body 3-D neuromechanical models can further relate these measurements to internal forces and body states otherwise difficult to access. For spiders, synchronized multi-view, high-speed measurements are needed to resolve rapid behavior and the resulting large-amplitude 3-D vibrations across the rapidly changing spider–web–prey system. Advances are also needed in methods for simultaneously measuring sensory responses in freely behaving spiders. Finally, experimentally validated, fast yet accurate dynamic models and simulations are needed to resolve vibrations during rapid behaviors and relate them to the underlying system mechanics. Across these systems, integrating multiple measurements at behaviorally relevant spatial and temporal scales with biologically validated neuromechanical models that estimate inaccessible mechanical variables will be particularly important for connecting interaction-generated mechanical information to its sensory and neuromuscular implementation.

## IMPLICATIONS FOR BIOLOGY AND ROBOTICS

For biology, these findings emphasize both deepening work in established model organisms and extending experimental approaches to diverse organisms and ecologically relevant environments. Established model systems have enabled remarkable progress in understanding biological function (Dallmann et al., 2023; Nishikawa et al., 2007; Tuthill and Wilson, 2016). Their experimental tractability often relies on simplified behavior and environments, which help isolate mechanisms but can obscure distributed, rapidly changing whole-body–environment interactions. Complementary experiments in more complex, ecologically relevant settings are therefore needed to determine how mechanisms identified under controlled conditions contribute to natural behavior (Dallmann et al., 2023; Ijspeert and Daley, 2023).

The diverse organisms and environments here illustrate that biological functions may emerge from mechanical interactions difficult to observe, measure, and manipulate. Robotic physical modeling can complement animal experiments, physiological measurements, and theoretical and computational modeling by reproducing salient interactions in controlled, instrumented systems. This enables systematic identification and manipulation of mechanical information before the relevant sensory receptors, neural circuits, and musculoskeletal processes are fully known. They can help generate focused, testable hypotheses about biological sensing and control in complex terrain. Continued development of measurements for freely behaving animals and complementary theoretical and computational models can help determine biological implementation for mechanosensing-guided whole-body behavior in physically complex environments.

For robotics, whole-body-generated mechanical interaction and feedback can be crucial for robust behavior in complex environments. Many current robotic approaches use vision-based environmental representations and data-driven learning (Ray, 2026), increasingly including learned world models (Hou et al., 2026). These approaches are powerful, but vision can be delayed or occluded in complex environments and does not directly measure rapidly changing local mechanical interaction variables such as contact forces, loading, and vibrations that determine whether a robot can move or otherwise function successfully (Chung et al., 2023; Ray, 2026; Roy et al., 2021). Data-driven approaches can also be data-hungry, particularly because collecting physical interaction data is costly and sample efficiency remains a major challenge in robot learning (Goldberg, 2025; Ray, 2026; Roy et al., 2021; Tang et al., 2025). The lack of parallelized, distributed, hierarchical mechanosensory processing may also contribute to the much higher closed-loop control bandwidth in robots than in animals (Dallmann et al., 2023). In addition, many current systems occupy a limited hardware design space and are developed for tasks and environments simpler than the diverse terrains in which animals operate. Even state-of-the-art terrestrial legged robots often remain less agile and robust than animals, especially in irregular or deformable terrains (Burden et al., 2024; Dallmann et al., 2023).

The three case studies show how interaction-generated mechanical information can support task-level behavior. More pervasive and elaborate use of mechanical sensing and feedback can complement vision, inertial sensing, mapping, simulation, and learned control (Ray, 2026; Roy et al., 2021). Such feedback can allow robots to react to unmodeled terrain features without fully reconstructing the environment (Wang et al., 2020), while mapping and navigation become more challenging with increasing environmental complexity and clutter (Chung et al., 2023).

More generally, robotic physical modeling can reduce apparently high-dimensional whole-body interaction problems to a smaller set of mechanically informative variables. This can facilitate robot designs that exploit morphological and environmental intelligence, rather than relying exclusively on environmental reconstruction and software control (Roy et al., 2021). Progress will require better integration of distributed mechanosensors, mechanosensing-informed control of physical interaction, and diverse task- and environment-specific hardware. In particular, morphology, materials, sensing, and actuation should be designed together to leverage the physical interactions most relevant to the task environment (Roy et al., 2021). Such integration will be important for advancing robots toward the agility and robustness that animals achieve in the diverse and complex real world (Burden et al., 2024).

## ACKNOWLEDGEMENTS

I thank my lab members for conducting the research reviewed here, especially Yaqing Wang, Eugene Lin, Qiyuan Fu, and Ratan Othayoth, who did the bulk of the work, and Sean Gart, Qihan Xuan, George Thoms, Jonathan Mi, Tommy Mitchel, Yuanfeng Han, Yulong Wang, Zhiyi Ren, Daniel Hsu, de la Tijera Obert, Evains Francois, Bokun Zheng, Ling Xu, Divya Ramesh, Yishun Zhou, and Luke Moon, who contributed. I also thank our collaborators, including Hsin-Yi Hung, Andrew Gordus, Henry Astley, Nathan Brown, Jochen Mueller, Jin-Seob Kim, and Greg Chirikjian, for their contributions. I thank Bob Full, Dan Goldman, Ron Fearing, Shai Revzen, Jean-Michel Mongeau, Simon Sponberg, Noah Cowan, and Beth Mortimer for helpful discussions. I thank Monica Daley, Matt McHenry, Auke Ijspeert, and Megan Leftwich for inviting me to present at the 2026 JEB Symposium, which led to this review paper. I thank the editor and three anonymous reviewers for helpful suggestions.

## COMPETING INTERESTS

The author declares no competing interests.

## FUNDING

The research reviewed in this paper was supported by an Arnold and Mabel Beckman Foundation Beckman Young Investigator award, a Burroughs Wellcome Fund Career Award at the Scientific Interface, an Army Research Office Young Investigator Program award, an NSF grant jointly supported by the Physics of Living Systems and Dynamics, Control, and System Diagnostics programs, and The Johns Hopkins University Whiting School of Engineering start-up funds. I thank Journal of Experimental Biology for travel support to present at the 2026 JEB Symposium, which led to this review paper.

## DATA AVAILABILITY

This is a review paper, and no new data are reported. All data discussed here are available in the cited research papers.

## USE OF AI TOOLS

During the preparation of this review, I used ChatGPT for brainstorming the organization of ideas and generating alternative phrasings of sentences, and I reviewed and edited the resulting content.